\documentstyle[aps,prl,twocolumn]{revtex}
\input epsf
\begin{document}
\title{Finger Competition and Formation of a Single Saffman-Taylor
Finger without Surface Tension: An Exact Result}

\author{Mark Mineev-Weinstein and Oleg Kupervasser}

\address{Theoretical Division, MS-B213, Los Alamos National Laboratory\\
Los Alamos, New Mexico 87545, USA\\Department of
Chemical Physics, \\ Weizmann Institute of Science, Rehovot 76100, Israel}

\maketitle
\begin{abstract}
We study the exact non-singular zero-surface tension solutions of the
Saffman-Taylor problem [Mineev-Weinstein and Dawson, Phys.~Rev. {\bf E 50},
R24 (1994); Dawson and Mineev-Weinstein, Physica {\bf D 73}, 373 (1994)]
for all times.
We show that all moving logarithmic singularities $a_k(t)$
in the complex plane $\omega = e^{i\phi}$, where $\phi$ is
the stream function,
are repelled from the origin, attracted to the unit circle and eventually
coalesce. This pole evolution describes essentially all the dynamical
features
of viscous fingering in the Hele-Shaw cell observed by Saffman and Taylor
[Proc. R. Soc. {\bf A 245}, 312 (1958)], namely tip-splitting, multi-finger
competition, inverse cascade, and subsequent formation of a single
Saffman-Taylor finger.
\end{abstract}
{\it Saffman-Taylor experiment.} In 1958 Saffman and Taylor [1] studied
the displacement of oil by water in a narrow gap between two parallel plates (the
Hele-Shaw cell [2]). The three upper pictures in Fig.1 show
three consecutive stages in their observations: initial, intermediate, and
asymptotic.
The small ripples in the left-most picture are caused by an instability
of the initially almost planar oil/water interface. We call this
initial stage {\it perturbative} since the deviation from a flat front is
small, so the dynamics can be examined by perturbation theory [3]. Later, at
the intermediate stage shown in the center picture, many fingers are
formed along the moving interface. They compete, some are screened and then
stagnate, some continue to advance and fatten, while new ones continue to
appear because of tip-splitting. We call this second (intermediate) stage
{\it non-perturbative}, since the deviation of the front from a straight line
is no
longer small and thus it cannot be studied perturbatively. Eventually, in the
final {\it asymptotic} stage (the right picture), a single uniformly
advancing finger dominates after all others are suppressed in the
competition described above.  This single finger occupies one half of the
Hele-Shaw cell width $w$ and is described by the formula
\begin{equation}
x = 2t + {\rm log}\,( \cos y)
\end{equation}
where $w = 2\pi$ in our scaled units; the longitudinal extent,
that is the $x$-direction, is taken to be infinitely large;
and velocities of both fluids are chosen to be 1 when
$|x|~\rightarrow~\infty$. Throughout the last 40 years after [1] was
published, this experiment has been reproduced many times, both in laboratories
and by numerical calculations. (See [4] and references therein). The process
above was found to be universal and independent of the specific liquids
(oil and water in [1]) provided the fluids are immiscible and
incompressible and a driving fluid is much less viscous than a driven one.

{\it A mathematical formulation} in the absence of
surface tension and in scaled units is
\begin{eqnarray}
\nabla^2 p = 0\quad {\rm (in~the\,oil~domain)} \nonumber \\
p =  - x\quad {\rm if}~~x\,\to +\infty \, \quad {\rm
(oil~pushed~to~the~right)} \nonumber \\
p = 0 \quad {\rm(at~ the~ oil/water~interface)} \label{eq2}
 \\
\partial_n p = 0 \quad {\rm for}~~y = \pm\pi \,\quad
\rm{(at~the~channel~walls)} \nonumber \\
V_n =- \partial_n p \quad \rm{(at~the~oil/water~interface) }
\nonumber
\end{eqnarray}
where $V_n$ is the normal velocity of the interface, $p$ is pressure, and
$\partial_n$ is the normal derivative.
From (2) follows the Laplacian growth equation (LGE)
(see [5] and references therein), which can be written compactly as:
\begin{equation}
\label{eq5}
{\rm Im}(\bar z_t z_{\phi}) = 1\,\,,
\end{equation}
for the moving front $z(t, \phi) = x(t, \phi) + iy(t, \phi)$,
parameterized by $\phi \in [0, 2\pi]$. Here the bar denotes the
complex conjugate; $z_t$ and $z_{\phi}$ are partial derivatives;
and the map $\phi \rightarrow z(t, \phi)$ is conformal for Im
$\phi \leq 0$.

{\it Derivation of the} LGE (3) from (2) follows.
Because of the identity $V_n = {\rm Im}(\bar z_t z_l)$, where $l$
is an arclength along the front, and of the Cauchy-Riemann relation,
$- \partial_n p~=~\partial_l\phi$ for pressure $p$ and stream function $\phi$,
where $\partial_l$ is the tangential derivative, the last equation in the
system (2) can be rewritten as
$${\rm Im}(\bar z_t z_l) = \partial_l\phi$$
which is equivalent to (3); ($\partial_l\phi = -\partial_n\,p \neq 0$, since
$p=0$ along the front and because of the maximum principle for harmonic
functions). It then follows that the variable $\phi$ in (3) is
the stream function. The map $\phi \rightarrow z(t, \phi)$ is conformal for Im
$\phi \leq 0$, since the complex velocity potential
$W(z) = -p(x,y) + i\phi(x,y)$ is analytic in the oil domain.

{\it Previous studies} have focused primarily
on the selection of the asymptotic finger width $\lambda$
from the continuous family, found in [1]:
\begin{equation}
x = \frac{t}{\lambda} + 2(1-\lambda)\,{\rm log} (\cos\frac{ y}{2\lambda})
\end{equation}
where the finger moves to the right with velocity $1/\lambda$, and
the finger's dimensionless width $\lambda$, measured in units of the channel
width, can be any
number between $0$ and $1$. (It is not difficult to see that (4) is the
traveling-wave solution of (3)). These selection studies focused on surface
tension as the critical factor, which determines the finger width  $\lambda$
[4,6]. (When $\lambda = 1/2$, (4) becomes (1)). Some studies (mostly numerical
and semi-analytical) also addressed the intermediate (non-perturbative) period
of the process [7], and again, as in the case of width selection, all the
main features observed during this stage were attributed to non-zero surface
tension. We believe that this attitude was formed because almost all exact
solutions known at that time [8]\footnote{Except for a few nonsingular examples
[9], limited however to a specific symmetry, and hence non-generic.} for
zero surface tension exhibited finite-time singularities (cusps). The
inevitability of cusps was related to the ill-posedness of the problem in the
absence of surface tension [10]. Recently however the selection of $\lambda =
1/2$ was solved [11] {\it in the absence of surface tension} using the exact
time-dependent $N$-logarithmic solutions of the LGE (3), which remain {\it
non-singular} and analytic {\it for all time} (no cusps)[5].

These {\it N-logarithmic solutions} have the form [5]:
\begin{equation}\nonumber
z(t, \phi) = \tau(t) +i\mu \phi + \sum_{k=1}^{N}\alpha_k
\log(e^{i\phi} - a_k(t)) \,,
\end{equation}
where $\alpha_k = {\rm const}$, $|a_k| < 1$,
$\mu = 1 - \sum_{k=1}^{N}\alpha_k$, and $\mu \in [-1,1]$.
With some constraints on $\{\alpha_k\}$, which are not specified here (see [12]
for details), these
solutions remain regular and do not exhibit finite-time singularities [5],
unlike all other known time-dependent solutions of (3).
The time dependence of $a_k(t)$ and $\tau(t)$ is given by
\begin{eqnarray}
&&\beta_k = \tau - \log \bar a_k + \sum_{l=1}^{N}\alpha_l\log(1 -
\bar a_k a_l) = \rm{const} \\
&&{\rm{and}}\,\,\,t + C = \tau  + \frac{1}{2}\sum_{k=1}^N \sum_{l=1}^N \alpha_k
\bar \alpha_l \log (1-a_k \bar  a_l)
\end{eqnarray}
where $C$ and $\beta_k$ ($k = 1, 2, ..., N$) are constants in time.
The validity of (6) and (7) can be easily verified after substitution
of (5) into (3) [13]. The solutions (5) are dense in the space of analytic
curves [14], and all singularities of $z_{\phi}$ are {\it simple} poles
$a_k(t)$, while all other known solutions of (3) [8] have multiple
poles. Thus (5) are the most natural and generic exact solutions of (3)
with a finite number of moving singularities.

{\it The geometrical interpretation of} (5) is of great help (see Figure 2):
The constant of motion $\beta_k - \alpha_k
\log 2$ is also the complex coordinate of the stagnation point at the vertex
of the $k^{th}$ ``groove'' with parallel walls. This groove originates during
the front evolution; $\pi|\alpha_k|$ is the width of this groove, and
arg($\alpha_k$) is the angle between the axis of symmetry of the groove
and the horizontal axis. This interpretation of solutions (5) is in
excellent agreement with all known low surface tension experiments; and we
claim that all stages of this process are faithfully described by (5) (to be
more exact - appropriate subset of (5) [15]).

{\it The goal of this paper} is to show how the exact solution (5)
describes essentially all stages of the Saffman-Taylor process,
but especially formation of a single Saffman-Taylor finger.
Incidentally, we also show how the zero-surface-tension solution (5) can
reproduce the following dynamical phenomena observed in this process and
attributed
traditionally to surface tension: tip-splitting, side-branching,
screening of retarded and coarsening of advanced fingers, and formation of a
single finger after finger competition.

To accomplish our goal we will
study the dynamical system of the poles $a_k(t)$ defined by (6-7) on the
mathematical plane $\omega = e^{i\varphi}$. As mentioned above, the dynamics
of poles
$a_k$ provides, through (5), a complete description of these phenomena in the
physical plane $z=x+iy$. In particular, the coalescence of all poles, which we
prove below, explains why only a {\it single finger} survives in the long-time
asymptotics.
The pictures at the bottom of Fig.1 show typical distributions of singularities
$a_k$ inside the unit circle associated with all three stages of this process.
\begin{figure}

\epsfxsize=9cm
\qquad
\epsffile{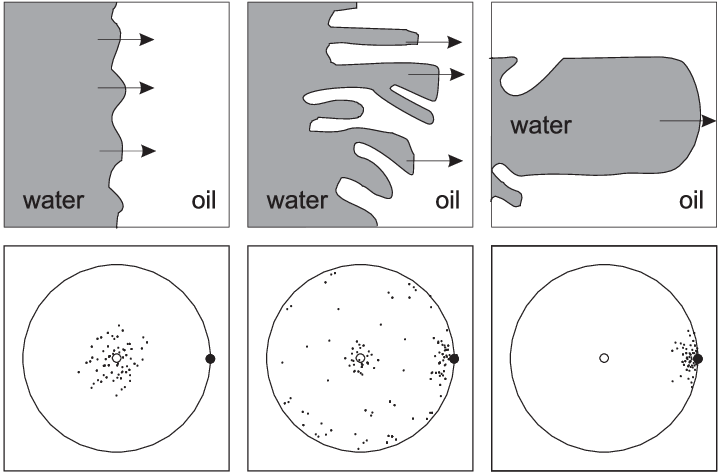}

\smallskip
\caption{\label{100}Three consecutive stages of viscous fingering
in the Hele-Shaw cell: initial (left), intermediate (center), and asymptotic
(right). The physical plane $z$ is shown in the upper pictures, while
the lower pictures depict a distribution of moving poles $a_k(t)$ in the
unit circle $|\omega|<1$ on the mathematical plane $\omega$. The open
circle indicates the repeller, $\omega = 0$, while the solid circle indicates
the attractor,
$\omega = 1$, of poles whose dynamics is given by (6-7).}

\end{figure}

{\it Initial stage: disordering begins, the repeller} at $\omega = 0$.
For a front that is almost flat initially, all singularities are near the
origin:
$|a_k|\,\ll\,1$ (see the left of Fig.1). Indeed, in this case the front is
almost planar: $z(0,\phi) \cong \tau + i\phi$. Then the singularities $a_k$
start
to move away from the origin. One can easily see from (6) that, if
$|a_k|\,\ll\,1$,
$|a_k| \cong e^{\tau}$. This exponentially fast repulsion from the origin,
holds for every pole $a_k$, if $|a_k|\,\ll\,1$,
regardless of the positions of the other poles. It means that $\omega = 0$ {\it
is the repeller} for the dynamical system of $a_k(t)$ specified by (6) and (7).
As one can easily see, $a_k=0$ is an unstable equilibrium solution of the
dynamical system (6-7). (The corresponding $\beta_k$ equals infinity.) The
repulsion from zero manifests itself in the front instability, which always
occurs when a less viscous fluid pushes a more viscous one. Indeed, the closer
$a_k$ comes to the unit circle, the more complex the interface
becomes.

Because of the exponentially fast instability, the system quickly
enters into the non-perturbative stage: convex parts of the interface
accelerate, while concave parts decelerate, retard, eventually stagnate and
start to form grooves near their most retarded points
(stagnation points). The tendency of moving fronts to form these grooves
with almost parallel walls has been observed in numerous laboratory and
computational experiments [4,6,7].
The dynamics of formation of these grooves coincides with the dynamics of the
solution (5), given by (6) and (7), near the points $z(t,\phi) \cong \beta_k
- \alpha_k \log 2$
and is fully compatible with the geometrical interpretation of (5) given above.
It is unknown, though, why these grooves have parallel (or almost parallel)
walls.

{\it Early intermediate stage: disordering increases.}
The formation of grooves near stagnation points explains such frequently
observed dynamical features as tip-splitting and side-branching of the moving
interface. Each such event makes the moving interface more complex:
it increases the effective number of grooves by one, and consequently
the total number of grooves grows in time.  (One can equally say that the
number of fingers grows in time, since a finger, as we traditionally call it,
is an element of the advancing front between two adjacent grooves. Note that while
grooves have parallel walls, fingers, as a rule, do not.) In short,
disordering and complexity of the front increases during this stage (the center
of Fig.1).

As shown in [5], the only long time asymptotics for each $a_k(t)$ consistent
with (6) is $|a_k| \rightarrow 1$. This was shown assuming that Re$\,\alpha_k
\geq 0$ for each $k = 1,2,...,N$. This assumption was taken in [5] as a working
criterion for the solutions (5) to be non-singular. Recent studies showed that
this criterion is too simple [12], but for many scenarios it works
successfully. Since (6) and (7) require all singularities $a_k(t)$ to
move toward the unit circle (see details in [5]), they eventually enter its
narrow vicinity and accumulate there. The $a_k$'s continue
to approach the unit circumference $|\omega|=1$ exponentially slowly, namely as
$1-|a_k| = c_k \exp{(-\tau/{\rm Re}\, \alpha_k)}$, when $c_k > 0$, Re $\alpha_k
> 0$, and when the poles $a_k$ are not too close to each other.
Only when $a_k$ approaches the unit circle does the
groove near the stagnation point $\beta_k - \alpha_k \log 2$ develop. In
this sense the geometrical interpretation of the constants $\alpha_k, \beta_k$
given above is applicable only in intermediate asymptotics, that is when
$|a_k|$ approaches one.

{\it Late intermediate stage: ordering begins.  ``The law of nonlinear
superposition'' for groove merging}. It was indicated and discussed in [5,16]
that when two poles $a_k$ are close to the unit circle, they are mutually
attracted along the angular direction, cannot pass each other and must merge.
This merging is complete only when $\tau \rightarrow \infty$, while for finite
$\tau$, the distance between merged poles is, as one can see from (6), of the
order of $e^{-b\tau}$, where $b>0$.  As two poles merge, say $a_1$ and $a_2$,
we equate them, so the total number of logarithmic terms in the sum in (5)
decreases by one, and a new logarithmic term in this sum forms from the two
terms labeled by $1$ and $2$. The coefficient of this new term clearly is
$\alpha_1 + \alpha_2$. In the physical plane this corresponds to the merging of
two grooves into a single one. The width and the angle (to the horizontal axis)
of the newly formed groove are $\pi|\alpha_1 + \alpha_2|$ and $\arg(\alpha_1 +
\alpha_2)$ respectively, while the widths and angles of two grooves before
their interaction were $\pi|\alpha_1|$, $\pi|\alpha_2|$ and $\arg \alpha_1$,
$\arg \alpha_2$ respectively. An example is shown in Fig.2.

\begin{figure}
\epsfxsize=5cm
\qquad \epsffile{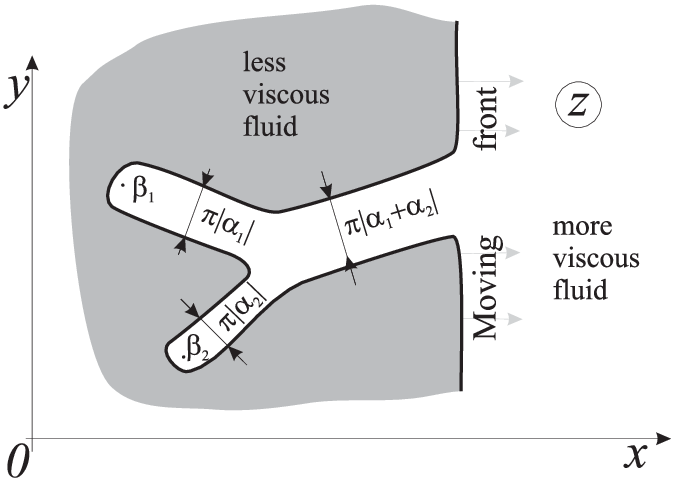}

\smallskip
\caption{\label{200}Geometrical interpretation of the complex
constants of motion $\alpha_k$ and $\beta_k$; $k=1, ..., N$.}
\end{figure}

This ``law of non-linear superposition'' of grooves governs the interaction
between these grooves and determines all the dynamics of groove coalescence.
Each interaction of this kind leaves one of interacting fingers behind,
while the others continue to advance and fatten, thus ``winning the
competition''. Two adjacent fingers do not interact only if they are strictly
parallel. Later we will show that this parallel configuration is linearly
unstable. As a result of this merging the total number of advancing fingers is
decreased by one. (Or, what is the same, the total number of effective grooves
decreases by one after each pair coalescence). In this fashion, the solutions
(5) describe the phenomenon of finger competition (or groove coalescence).
This ordering process eventually dominates the disordering due to tip-splitting
and side-branching (which is just the birth and development of new grooves).
So, in the long-time asymptotics only a single finger is left to propagate
(Fig.1 upper-right).

Physically, the merging
of grooves is an inverse cascade, an ordering process that
dominates after saturation. This self-organizing behavior of an unstable front
is not exceptional: it occurs in several unstable nonlinear
processes in confined geometries [17]. It is the walls of the
box the system is contained in, which stabilize growth. This happens when the
transverse correlation length, initially small, becomes comparable with the
lateral width of the box [18]. Then in all these processes a universal pattern
appears (different for different processes). In the dynamics described by (2),
this pattern is the single finger, observed by Saffman and Taylor in [1].

The rest of this paper derives the coalescence of {\it all} singularities
$a_k$ at a single point on the unit circle
(for periodic boundary conditions at the walls) or at two opposite
points on the unit circle $a_k \rightarrow \pm 1$ (for no-flux boundary
conditions at the walls). In the physical plane, this means that, for generic
initial data in the class of non-singular solutions (5), the asymptotics $t
\rightarrow \infty$ is indeed a single uniformly advancing finger in accordance
with experimental observations. Then replacing every $a_k$ by 1 and $\tau$ by
$Vt$ in (5) when $t \rightarrow \infty$, we obtain
$$z(t, \varphi) = Vt + \mu i \varphi + (1-\mu) \log (e^{i\varphi} - 1)$$
which coincides with (4) when $V = 1/\lambda$ and $\mu = 2\lambda-1$ and
thus describes a single uniformly moving finger, as expected. (To
select $\lambda = 1/2$ within the class (5), it is sufficient to move
the pole infinitesimally away from the unstable equilibrium that is zero, as
was shown in [11].)

{\it The variable} $\tau$ {\it is a scaled time}. First we prove that $\tau$
in (5-7) is but a scaled time, and it will be
treated as such below. First, we note that $d\tau/dt$ is always finite, since
we consider only {\it non-singular} solutions (5). Then we see that always
$\tau >t$. To prove this, we expand every logarithm in the RHS of (7) as an
infinite power series. We obtain
$$t  = \tau - \frac{1}{2} \sum_{m=1}^{\infty} \frac{|\sum_{k=1}^N
\alpha_k a_k^m|^2}{m}$$
(setting $C=0$ in (7).)
Thus $\tau >t$. Then if $t\,\rightarrow\,\infty$,
so does $\tau$.
It follows from (6) and (7) that
\begin{equation}
t + C'  =  \frac{1+\mu}{2}\tau + \frac{1}{2}\sum_{k=1}^{N} \alpha_k  \log(a_k)
\end{equation}
where $C'$ is a constant. The only way to compensate the divergent
positive term $\tau$ in the RHS of (6) is to admit that $|a_k|\,\rightarrow\,1$
for each $k=1, 2, ..., N$ [19].The only way to compensate the divergent
positive term $\tau$ in the RHS of (6) is to admit that $|a_k|\,\rightarrow\,1$
for each $k=1, 2, ..., N$ [19].
Therefore in the long-time limit the relation between $t$ and $\tau$ is
linear.

{\it Proof that all poles coalesce in the long-time limit.}
We want to investigate the asymptotic behavior of the poles $a_k$ when $\tau
\rightarrow \infty$. First, we introduce new variables $a_k = \exp\{-\eta_k +
i\Phi_k\}$ and  $\alpha_k = \alpha'_k + i\alpha''_k$. The condition $|a_k|
\rightarrow 1$ then simply means that $\eta_k \rightarrow 0+$. In order to
eliminate the divergent term $\log(1 - |a_k|^2)$ from (6) we multiply (6) by
$\bar \alpha_k$ and then extract its imaginary part. The result is
\begin{eqnarray}
&&{\rm Im}(\bar \alpha_k \beta_k) = \alpha'_k \Phi_k - \alpha''_k \tau \nonumber \\
&&+ \sum_{l \neq k}^N [(\alpha''_l \alpha'_k - \alpha''_k \alpha'_l)\log|1 - \bar a_k a_l| \nonumber \\
&&+ \,(\alpha'_k \alpha'_l + \alpha''_l \alpha''_k)\,{\rm arg}(1 -
\bar a_k a_l)].
\end{eqnarray}
(We neglected the small term, $-\alpha''_k \eta_k$).
Here we have the divergent term $- \alpha''_k \tau$. To compensate this
divergence
we need to assume that at least some poles coalesce. That is $N$ poles
$a_k$ are divided into $\tilde{N}$ groups of merging poles $a_k$ so that
$|\Phi_k - \Phi_l| \rightarrow 0$ for all members of each group. After
summation of eqs. (9) over all poles within each of the $\tilde N$ groups, we
obtain
\begin{eqnarray}
&&C_m = \tilde \alpha'_m \tilde \Phi_m  - \tilde \alpha''_m \tau  \nonumber \\
&&+ \sum_{n \neq m}^{\tilde N} [(\tilde \alpha''_n \tilde \alpha'_m - \tilde
\alpha''_m \tilde \alpha'_n)\log|1 - \bar {\tilde a_m} \tilde a_n| \nonumber
\\
&&+ (\tilde \alpha'_m \tilde \alpha'_n + \tilde \alpha''_m \tilde \alpha''_n)\,
{\rm arg}(1 - \bar {\tilde a_m} \tilde a_n)]\,\,.
\end{eqnarray}
Here $C_m$ are constants, $\tilde \Phi_m = \arg  \tilde a_m , \, \tilde
\alpha'_m = \sum_{k=1}^{n_m}\alpha'_k$,\,
$\tilde \alpha''_m = \sum_{k=1}^{n_m}\alpha''_k$, $n_m$ is the number of poles
in the $m$-th group, and $1\, \leq \, m\, \leq \,\tilde N$. Also
because of the partial merging, all poles within each group labeled by $m$
are
equated to $\tilde a_m$ with an exponentially high accuracy. Replacing (with
the same accuracy) $|\tilde a_m|$ by 1, and assuming no merging between
different $\tilde a_m$'s, which means
$$|\Phi_k - \Phi_l| \gg \eta_k + \eta_l$$
where $k$ and $l$ are from different groups,
we rewrite (10) as
\begin{eqnarray}
&&C'_m = \tilde \alpha'_m \tilde \Phi_m - \tilde \alpha''_m \tau  \nonumber \\
&&+ \sum_{n \neq m}^{\tilde N}[(\tilde \alpha''_n \tilde \alpha'_m -\tilde
\alpha''_m \tilde \alpha'_n)\log|\sin(\frac{\tilde \Phi_m - \tilde
\Phi_n}{2})|
\nonumber \\
&&+ (\tilde \alpha'_m \tilde \alpha'_n  + \tilde \alpha''_n \tilde \alpha''_m)
\frac{\tilde \Phi_n - \tilde \Phi_m}{2}],
\end{eqnarray}
where $C'_m$ are new constants. From this equation we conclude that

(i) Since $\log|\sin(\frac{\tilde \Phi_m - \tilde \Phi_n}{2})| \rightarrow
-\infty$ for $|\tilde \Phi_m - \tilde \Phi_n|\, \rightarrow 0, 2\pi$,
the poles cannot pass each other.

(ii) After summation over all $m=1, ..., \tilde N$, we see that $\sum_m \tilde
\alpha'_m \tilde \Phi_m = {\rm const}$.

(iii) From (i) it follows that once $0 < \tilde \Phi_m - \tilde \Phi_n <
2\pi$  in the long-time asymptotics it will hold for all further times.

(iv) From (ii) and (iii) it follows that $|\tilde \Phi_m| \rightarrow \infty$
is impossible.

(v) From (iii) and (iv) it follows that there is no way to compensate the
divergent term  $- \tilde \alpha''_m \tau$ in (11), except to admit that
$\tilde \alpha''_m \, = \, 0$ for each $l=1, ..., \tilde N$.

Finally we conclude that
$$\tilde \alpha''_m = \sum_{j=1}^{n_m}\,\alpha''_j = 0, \,\,\,\, \dot
{\tilde \Phi_m} = 0, \,\,\,\,
\tilde \Phi_m \neq \tilde \Phi_n \, ({\rm for}\, m \neq n) $$
To find the asymptotic motion of poles within each group containing $n_m$ poles
we differentiate the equations (7) keeping only leading (divergent) terms:
$$\dot \tau \,+\,\sum_{l=1}^{n_m} \alpha_l \frac{\dot \eta_l + \dot
\eta_k + i(\dot \Phi_k - \dot \Phi_l)}{\eta_l + \eta_k + i(\Phi_k - \Phi_l)}
= 0$$
The solution of these equations is
\begin{eqnarray}
&&\eta_k = \eta_k^0 e^{-\tau/\tilde \alpha'_m}  \nonumber \\
&&\Phi_k - \Phi_l = (\Phi_k^0 - \Phi_l^0) e^{-\tau/\tilde \alpha'_m}  \nonumber
\end{eqnarray}
provided that
\begin{equation}
\tilde \alpha''_m = \sum_{k=1}^{n_m}\,\alpha''_k = 0.
\end{equation}
{\it Instability of configurations with two or more parallel fingers.}
Therefore, the only case when merging will not be complete occurs when two or
more groups of sums of $\alpha_k$'s, that are $\tilde \alpha_m$'s, are real.
This corresponds to the case (see the geometrical interpretation of $\alpha$'s
above) when all newly formed grooves characterized by $\tilde \alpha_m$'s are parallel to
each other (and to the horizontal axis). In this case instead of a single finger
in the asymptotics, we would have several parallel fingers with grooves between
them having widths of
$\pi\,|\tilde \alpha_m|$.  But this configuration is unstable:
In order to verify this, let us perturb slightly the constants $\alpha''_k$ so
that (12) is violated for every $m = 1, \ldots , \tilde N$, or we may merely
add new logarithmic terms to the sum in (6). Once we destroy
(even slightly) the parallelism of the newly formed grooves that are
partial sums of $\alpha_k$'s, then in accordance with the analysis above we
have only one group of poles which satisfy (12). And this group
contains all poles, so $\tilde N=1$. Hence we have total coalescence of all
poles $a_k$. In the physical plane this means merging into a single groove
(or, what is the same, a front forms a single finger).

The unique solution for $t \rightarrow \infty$ for each pole is
\begin{eqnarray}
&&\eta_k = \eta_k^0 e^{-2t/(1-\mu^2)}  \nonumber \\
&&\Phi_k = \Phi_k^0 e^{-2t/(1-\mu^2)}  \nonumber \\
&&1-\mu^2 > 0
\end{eqnarray}
Here we used from (8) that $(1+\mu)\tau = 2t$ for $t \rightarrow \infty$.
Thus for periodic boundary conditions all poles $a_k$ coalesce at a single point on the unit circle
(chosen to be $\omega = 1$ in Fig.1). In the case of no-flux boundary conditions\footnote{expressed in (2) with $\partial_n p = 0$ at the channel walls},
which impose double periodicity\footnote{The channel width becomes $\pi$
in order to preserve the solutions (5).} with reflectional symmetry on this
problem
[11] (see also [20]), there are clearly two groups of poles mutually complex
conjugate to each other. Because of this (mirror) symmetry, the attractors
are real, and there are clearly two of them, $\omega = \pm 1$, in this
case, instead of one attractor as for periodic boundary conditions.

{\it Proof of total coalescence without any formulae}. After the detailed
analysis (10)-(13) leading to the necessity of total coalescence, we will
show how to prove the main result of this paper without a single formula, by
using only the geometrical interpretation of (5) and the ``law of non-linear
superposition'' given above: Any two adjacent grooves whose axes of symmetry
are not strictly parallel (which is clearly the generic case) will eventually
coalesce. This will happen where their axes of symmetry cross. In accordance
with Euclid, two non-parallel straight lines on the plane necessarily cross,
and since each crossing means the coalescence of two grooves, eventually all
grooves coalesce. At first glance, this proof is valid only for those
grooves whose axes of symmetry cross in advance of the moving front, and
invalid when they cross behind it. But the proof still works because of the
mirror symmetry imposed by no-flux boundary conditions (see [11]). This
implies that if the middle lines of two adjacent grooves cross behind the
front, then one middle line and the mirror image of the other one necessarily
cross in front of the interface, so again coalescence is unavoidable.

{\it Summary of the asymptotic stage.} After a cascade of subsequent groove
crossings and coalescence, new grooves will be formed, described by $\tilde
\alpha_m$, which are sums of the $\alpha_k$'s of the original grooves
participating in
coalescence. New grooves will coalesce by the same rules as the original ones,
that is when their symmetry axes cross. Since the parallel configuration of two
or several grooves is unstable, eventually a single groove will be formed with
a width of $\pi(1-\mu)$. Since this is a real number, this final single groove
will be parallel to the Hele-Shaw channel's walls. Then the single finger (a
moving front between two grooves) will propagate along the Hele-Shaw cell in
accordance with observations. The finger width is $2\pi - \pi(1-\mu) =
\pi(1+\mu)$, and the relative finger width measured in units of Hele-Shaw cell
(chosen here to be $2\pi$) is $\lambda = \pi(1+\mu)/(2\pi) = (1+\mu)/2$.
Regarding the selection of $\lambda = 1/2$ from this family (that is, $\mu =0$)
we will just repeat here the essence of [11],
where two observations were made: that the origin is a point of unstable
equilibrium (the repeller) for singularities, and that all fingers with a width
different from one half correspond to the case when one of the singularities lies
at the origin. Once this singularity is perturbed (released from zero),
it moves with the rest of poles toward the attractor lying on the unit
circle [11]. Then, once there are no singularities at zero, the asymptotic
finger has a width $\lambda = 1/2$, in accordance with experiments [1,4].

{\it General summary.} The solutions (5) indeed explain all stages of front
dynamics in the Hele-Shaw cell in great detail. On the mathematical plane,
the singularities $a_k$ move away from the repeller at the origin, where most
of them are initially located, toward the unit circle which is the
intermediate attractor for the dynamical system (6,7). On the physical plane
this
is a process of disordering which occurs in agreement with the geometrical
interpretation, that is through the formation of grooves described by constants
$\{\alpha_k, \, \beta_k\}$, until the saturation time marks the end of the {\it
disordering} part of the process. Then the
system enters into the {\it ordering part}, when the number of competing
fingers eventually decreases and finally a single finger survives the
competition and persists to advance in the long-time asymptotic limit
(Fig.1 upper-right). On the mathematical plane it means that when the radial
component of poles' velocities becomes small near the unit circle, the angular
component brings pairs of singularities together, and they attract and
coalesce. Then groups of coalesced singularities move as units near the unit
circle and eventually merge together at a single point on the unit circle
($\omega = 1$ in Fig.1), which is the {\it only} attractor in periodic
boundary conditions. The no-flux boundary conditions imposing the reflectional
symmetry make {\it two} attractors at $\pm1$.

{\it Conclusion.}
We have shown that the front expressed by non-singular
solutions (5), which describe
Laplacian growth in the Hele-Shaw
cell {\it in the absence of surface tension}, forms a single finger uniformly
propagating along the Hele-Shaw cell asymptotically in time.
Essentially all the observed dynamical features of tip-splitting, finger
competition, screening, coarsening, and finally forming a single finger are
shown to be consequences of the solution (5) for a generic choice of initial
parameters.

{\it What is unclear.} Here we mention questions for further elucidation: We
have studied the asymptotics far beyond the farthest stagnation point
$\beta_k$, implying that for some reason perturbations of the front eventually
stop. What physically happens is that during the formation of the single
finger, the front gradually stabilizes and eventually is no longer vulnerable
to weak noise. Clearly, this stabilization cannot be justified without non-zero
surface tension. It is remarkable that the zero surface tension solution (5)
nevertheless works and explains all these phenomena including tip-splitting,
side-branching, finger competition, screening and coarsening [5], formation of
a single finger (the present paper), and the selection of its width of one
half [11], in accordance with numerous observations. Why does it work without
surface tension? And how does surface tension regulate population and location
of stagnation points and width of grooves? The answers to these questions are
not known today.

{\it Acknowledgement.} We gratefully acknowledge discussions with G.~D.~Doolen,
J. Pearson, and I. Procaccia. We also appreciate help of P. Vorobieff with
computer imaging and the financial support of the Department of Energy
programs at LANL.



\begin{references}
\bibitem{13} In [5], $t$ was mistakenly written instead of $\tau$ in the
right-hand side of the analog of (6). This does not change results of
[5], since $\tau$ can be treated as a scaled time, as shown in this paper.
\bibitem{19} This is obvious if $\alpha_k'\,>\,0$ for each $k$ [5] and less
obvious, but still true, if not [12].
\end{references}
\end{document}